\documentclass{article}

\usepackage{arxiv}

\usepackage[utf8]{inputenc} 
\usepackage[T1]{fontenc}    
\usepackage{hyperref}       
\usepackage{url}            
\usepackage{booktabs}       
\usepackage{amsfonts}       
\usepackage{nicefrac}       
\usepackage{microtype}      
\usepackage{lipsum}		
\usepackage{graphicx}
\usepackage[numbers]{natbib}
\usepackage{doi}
\usepackage{amsmath}
\usepackage{amssymb}
\usepackage{subcaption}
\usepackage{cleveref}
\usepackage{algorithm}
\usepackage{algpseudocode}
\usepackage{tikz}
\usepackage{dot2texi}

\usetikzlibrary{positioning, decorations.pathreplacing}

\crefname{algorithm}{algorithm}{algorithms}
\Crefname{algorithm}{Algorithm}{Algorithms}

\title{mRNA Folding Algorithms for Structure and Codon Optimization}

\author{ \href{https://orcid.org/0000-0001-9114-7339}{\includegraphics[scale=0.06]{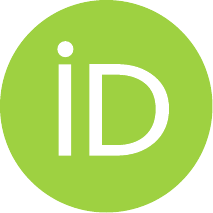}\hspace{1mm}Max Ward}\thanks{Co-corresponding authors} \\
	School of Physics, Mathematics, and Computing\\
	The University of Western Australia\\
	Perth, Australia\\
	\texttt{max.ward@uwa.edu.au} \\
	\And
	\href{https://orcid.org/0000-0002-0177-6789}{\includegraphics[scale=0.06]{orcid.pdf}\hspace{1mm}Mary Richardson} \\
	Moderna, Inc.,\\
	Cambridge, MA, USA \\
    \And
    Mihir Metkar\thanks{Co-corresponding authors} \\
	Moderna, Inc.,\\
	Cambridge, MA, USA \\
	\texttt{mihir.metkar@modernatx.com} \\
}

\renewcommand{\shorttitle}{mRNA Folding Algorithms for Structure and Codon Optimization}

\begin{document}
\maketitle

\begin{abstract}
    mRNA technology has revolutionized vaccine development, protein replacement therapies, and cancer immunotherapies, offering rapid production and precise control over sequence and efficacy. However, the inherent instability of mRNA poses significant challenges for drug storage and distribution, particularly in resource-limited regions. Co-optimizing RNA structure and codon choice has emerged as a promising strategy to enhance mRNA stability while preserving efficacy. Given the vast sequence and structure design space, specialized algorithms are essential to achieve these qualities. Recently, several effective algorithms have been developed to tackle this challenge that all use similar underlying principles. We call these specialized algorithms \textit{mRNA folding} algorithms as they generalize classical RNA folding algorithms. A comprehensive analysis of their underlying principles, performance, and limitations is lacking. This review aims to provide an in-depth understanding of these algorithms, identify opportunities for improvement, and benchmark existing software implementations in terms of scalability, correctness, and feature support.
\end{abstract}


\section{Introduction}\label{sec:intro}
The development of mRNA vaccines against Severe Acute Respiratory Syndrome Coronavirus 2 (SARS-CoV-2) has unequivocally demonstrated the potential of mRNA therapeutics to combat and control infectious diseases \cite{fang2022advances, hogan2022mrna}. Beyond vaccines, mRNA therapeutics are showing significant promise in early-phase clinical trials for cancer neoantigen vaccines, enzyme replacement therapies, and as a delivery platform for gene-editing enzymes \cite{shi2024progress}, paving the way for treatments targeting diverse medical conditions. As informational molecules, mRNAs encode the desired therapeutic protein directly within their sequence, offering unparalleled flexibility in design and production \cite{metkar2024tailor}. This adaptability positions mRNA as a versatile platform for addressing numerous therapeutic challenges.

Despite its advantages, the inherent instability of mRNA remains a significant barrier to its widespread use. mRNAs are highly prone to degradation by hydrolysis, necessitating ultracold storage and specialized supply chains to preserve in-vial stability \cite{uddin2021challenges}. These logistical hurdles disproportionately affect resource-poor regions, restricting access to mRNA-based medicines. Overcoming mRNA's storage and transport instability is therefore crucial to improving its global distribution, scalability, and equitable access.

One promising strategy to enhance mRNA stability is the co-optimization of RNA secondary structure and codon usage. RNA structure plays a pivotal role in determining susceptibility to hydrolytic degradation \cite{wayment2021theoretical}, while codon usage affects translational efficiency and protein expression levels \cite{metkar2024tailor, sharp1987codon}. However, for any given protein sequence, the number of possible mRNA sequences and their associated structures is astronomically large. For instance, SARs-CoV2 spike protein has $10^{632}$ possible nucleotide sequences with each having $\approx2.3^{3819}$ possible secondary structures \cite{metkar2024tailor}. As a result, computational algorithms have become indispensable for designing mRNA sequences that balance high in-vial stability with sufficient in-cell translation.

In recent years, a growing number of algorithms have been developed to address this multi-objective optimization problem \cite{zhang2023algorithm,wayment2021theoretical,leppek2022combinatorial, vostrosablin2024mrnaid, li2024codonbert}. This review focuses on a popular type of algorithm that we call ``mRNA folding'' algorithms. Although these have been popular recently \cite{zhang2023algorithm,gu2024derna}, the foundational concepts appeared earlier \cite{cohen2003natural, terai2016cdsfold}. The fundamental strategy mRNA folding algorithms employ is to extend standard RNA folding algorithms \cite{zuker1981optimal,zuker2003mfold,reuter2010rnastructure,lorenz2011viennarna} by incorporating mRNA-specific constraints such as codon usage biases \cite{sharp1987codon, zhang2023algorithm} to generate optimized mRNA sequences.

This review aims to introduce the fundamentals of mRNA folding algorithms, highlight gaps in research, and propose opportunities for future improvements. Additionally, we will provide a comprehensive comparison and benchmark of existing software packages that implement mRNA folding algorithms. By offering a foundational overview of this rapidly evolving subfield of mRNA therapeutics, this review aims to guide researchers in selecting and improving algorithms for the rational design of next-generation mRNA therapeutics. Other tools, such as RiboTree \cite{wayment2021theoretical}, incorporate mRNA folding algorithms in their sequence design process. However, since they primarily combine mRNA folding with heuristics, they will not be a focus in this review.

\subsection{An Overview of mRNA Folding Algorithms}

Existing mRNA folding algorithms are based on the dynamic programming method first introduced by Zuker and Stiegler in 1981 for RNA secondary structure prediction \cite{zuker1981optimal}. This algorithm forms the core of modern RNA structure prediction tools \cite{lorenz2011viennarna,reuter2010rnastructure,zuker2003mfold, zadeh2011nupack}. This algorithm and similar techniques are often called ``RNA folding'' algorithms. Extending the nomenclature, we define algorithms that leverage RNA folding principles for mRNAs design as ``\textit{mRNA folding}'' algorithms.

Currently, four mRNA folding algorithms are described  in the literature. The first, published by Cohen and Skiena \cite{cohen2003natural}, maximized mRNA structure. This was followed by CDSfold \cite{terai2016cdsfold}, which improved the algorithmic efficiency significantly and added several additional capabilities. Both methods have proven prescient as they predate the recent surge of interest in mRNA design. These algorithms modify the Zuker-Stiegler dynamic programming recursions to minimize free energy under codon constraints. Formally, let $\textsc{MFE}(\pi)$ calculate the Minimum Free Energy (MFE) (e.g., by using the Zuker-Stiegler algorithm), and let $\Psi$ be the set of valid protein-coding sequences, then they calculate $\operatorname{argmin}_{\pi \in \Psi}{\textsc{MFE}(\pi)}$. The intuition is that a lower MFE implies higher stability \cite{zur2012strong, leppek2022combinatorial,wayment2021theoretical}. The Cohen-Skiena method achieves this by adding codon conditions to the Zuker-Stiegler recursions. CDSfold instead uses a graph of valid codon sequences, which led to a significantly faster algorithm. Both methods have a notable limitation: they cannot simultaneously optimize stability (via MFE) while maintaining high translation efficiency (measured by Codon Adaptation Index (CAI) \cite{sharp1987codon}).

The next method was LinearDesign \cite{zhang2023algorithm}, which mitigated this limitation by co-optimizing for MFE and CAI. It also improved on CDSFold by incorporating a beam search heuristic, substantially increasing algorithmic speed at the moderate cost of a potentially approximate optimized mRNA.

LinearDesign balances the MFE and CAI weights using a mixing factor $\lambda$, defining the sequence-structure score as $\textsc{MFE}-\log(\textsc{CAI})\times \lambda$. However, this approach presents two challenges: first, if the user wants to target a specific CAI, they need to search for the right $\lambda$; second, those unsure about target CAI need to make an arbitrary choice. Another recent alternative, DERNA \cite{gu2024derna}, addressed this limitation by finding all Pareto optimal solutions for CAI and MFE, thus allowing users to find the best MFE for every possible CAI. However, DERNA has some drawbacks compared to LinearDesign and CDSfold. It is slower, even when not computing the Pareto optimal frontier. Gu \textit{et al.} reported a 6-hour maximum run time for DERNA on their benchmarks versus 19 minutes for LinearDesign \cite{gu2024derna}. This is because DERNA extends the older codon-condition-based approach from the Cohen-Skiena algorithm, rather than the faster graph-based approach introduced by CDSfold and extended by LinearDesign.

From a practitioners standpoint, the choice is between using LinearDesign, CDSfold, and DERNA. Cohen and Skiena's method is superseded by the newer approaches and lacks publicly available source code. The publicly available version of LinearDesign supports CAI as well as MFE optimization and offers speed at the tradeoff of using a heuristic (beam search). CDSfold, while more efficient than DERNA, supports only MFE optimization. DERNA supports both Pareto optimization for CAI and MFE, but uses a slower algorithm similar to the older Cohen and Skiena's method.

The details of how mRNA folding algorithms work is only partially available in the literature. CDSfold algorithm is fully explained \cite{terai2016cdsfold}, but it does not incorporate CAI. Only a simplified version of LinearDesign \cite{zhang2023algorithm} is explained that omits the full algorithm. DERNA \cite{gu2024derna} and the Cohen-Skiena algorithm \cite{cohen2003natural} are described in full, but use inefficient algorithms.

An explanation of the complete mRNA folding algorithm that efficiently incorporates CAI is not available in the existing literature. To address this gap, \Cref{sec:mrna_folding} this review provides a comprehensive explanation of how these mRNA folding algorithms are constructed, including full algorithmic details. We simplify and unify existing approaches by introducing a ``codon graph'' framework. Finally, \Cref{sec:benchmarks} presents benchmarks for existing mRNA folding software packages. These include performance and correctness comparisons. We begin by explaining foundational definitions and concepts.

\section{The mRNA Folding Problem}

An mRNA coding sequence (CDS) encodes a protein. A protein is defined by a sequence of amino acids $\alpha = \alpha_1, \alpha_2, \dots, \alpha_n$. Each amino acid is encoded by multiple synonymous codons \cite{sharp1987codon}. A codon is considered valid for a given amino acid if it belongs to the set of synonymous codons for that amino acid. A valid CDS for $\alpha$ is a sequence of codons where each codon is valid for the corresponding amino acid.

\subsection{Preliminary Definitions}
We start with fundamental definitions useful in describing RNA and mRNA folding algorithms.

\begin{figure}[h]
    \centering
    \includegraphics[width=1.0\textwidth]{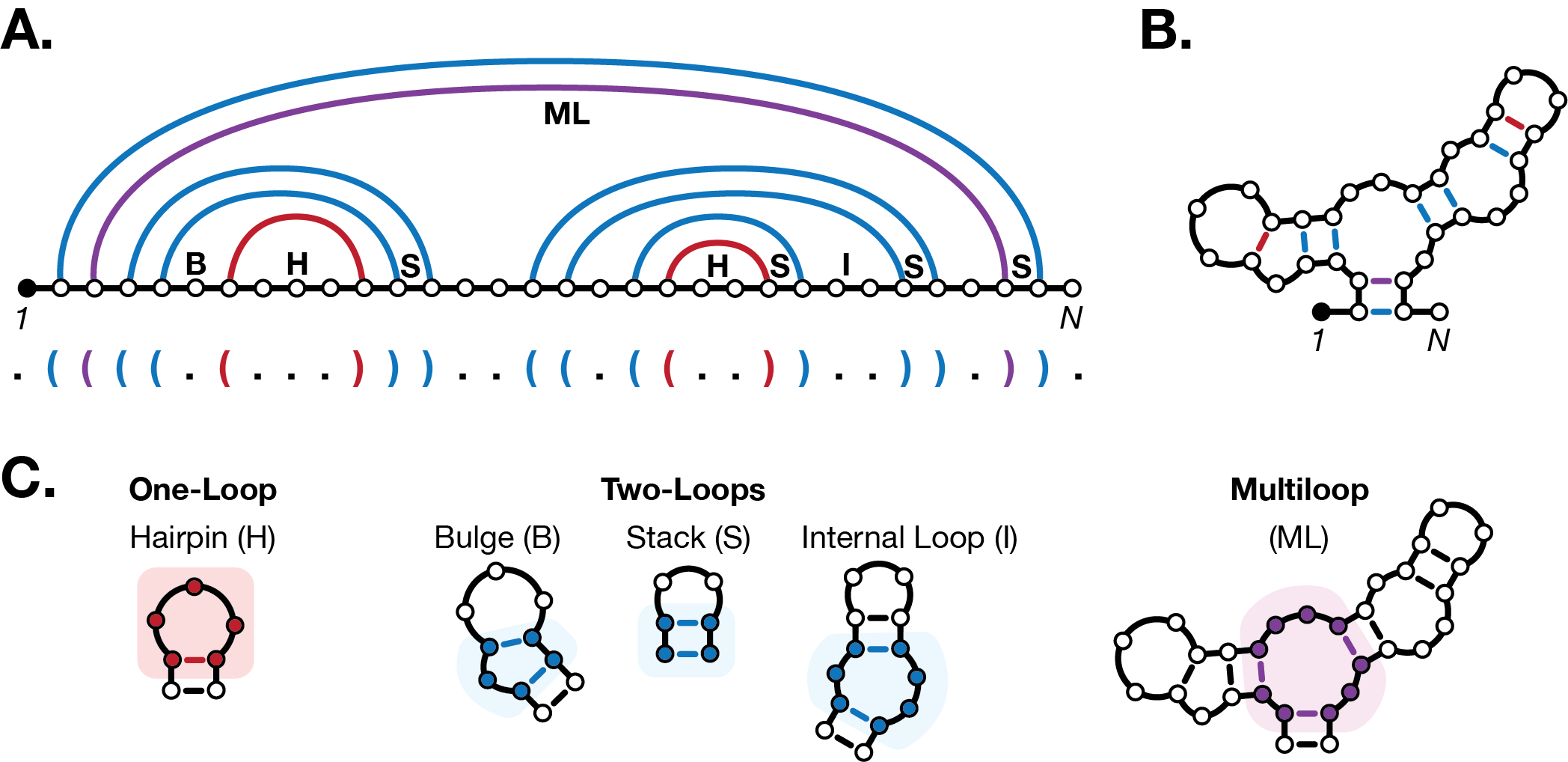}
    \caption{RNA secondary structure elements. (A) and (B) are different diagrammatic representations of the same example nested RNA structure. (A) is an RNA arc diagram, with the corresponding dot-bracket notation for the structure shown below. (B) is an RNA secondary structure diagram. (C) illustrates the three types of loops considered in RNA secondary structure prediction. Colors and labels in (A) and (B) correspond to the loop types in (C). }
    \label{fig:fig1}
\end{figure}

Given an RNA sequence $\pi$ we can define a set $S$ of valid structures. In the Zuker-Stiegler algorithm, we define a valid structure as a properly nested secondary structure (see \Cref{fig:fig1} panel A).

Formally, let $\pi$ represent an RNA sequence. An RNA is a sequence of nucleotides denoted by `A', `U', `G', and `C': $\pi \in \{\text{A}, \text{U}, \text{G}, \text{C}\}^\star$. A valid structure $s \in S$ is a set of pairs representing bonds between nucleotides. Only three nucleotide combinations can pair: \text{AU}, \text{GC}, \text{GU}. Note that these can pair in either orientation, e.g., \text{AU} and \text{UA} are both valid pairs.

A single nucleotide can be in at most one pair in a valid structure: $(i,j) \in s \implies (x,y) \ni s \text{ such that } (x,y) \neq (i,j) \text{ and } (x=i \text{ or } x=j \text{ or } y=i \text{ or } y=j)$. A valid structure contains no crossing pairs. Two pairs $(i, j), (k,l) \in s$ cross iff $i<k<j<l$ or $k<i<l<j$.

\subsection{The Objectives of mRNA Folding}

Early mRNA folding methods aimed to identify the coding sequence (CDS) that minimizes Minimum Free Energy (MFE), producing the most stable structure among all valid CDSs for a target protein \cite{terai2016cdsfold, cohen2003natural}.

The MFE structure of an RNA can be found using RNA folding algorithms, such as the Zuker-Stiegler algorithm. RNA folding algorithms require an energy function $E(s | \pi)$ that gives the free energy change for the sequence $\pi$ folding into the structure $s$. The goal of these algorithms is to compute the structure with the minimum free energy under $E$, with ties broken arbitrarily:

\begin{equation}\label{eq:rna-folding-mfe}
    \textsc{MFE}(\pi) = \operatorname{argmin}_{s} E(s | \pi)
\end{equation}


RNA folding algorithms predict the structure of a single RNA sequence. mRNA folding algorithms extend RNA folding to consider all coding sequences that could possibly encode the target protein. Early mRNA folding algorithms directly identify the sequence $\pi$ with the lowest MFE structure from the set of all valid coding sequences $\textsc{CDS}(\alpha)$ (\cite{terai2016cdsfold}, \cite{cohen2003natural}):

\begin{equation}\label{eq:mrna_fold_nocai}
    \textsc{Fold}(\alpha) = \operatorname{argmin}_{\pi \in \textsc{CDS}(\alpha)} \textsc{MFE}(\pi)
\end{equation}

However, the goal of mRNA folding is to find the optimal sequence-structure pair for the CDS, rather than just the optimal structure as in RNA folding. In addition to structural stability, codon usage is an important factor \cite{metkar2024tailor}. The optimality of a codon sequence is often calculated using a metric called Codon Adaptation Index (CAI) which measures adaptation of a sequence to the host organism \cite{sharp1987codon}):

\begin{equation}\label{eq:cai}
    \textsc{CAI} = \sqrt[^{|\alpha|}]{\prod_{i}{\frac{f_i}{\max(f_j)}}}
\end{equation}

The CAI is the geometric mean of codon scores derived from a set of highly expressed genes in the target host. CAI values range from 0 to 1, where 1 indicates perfect adaptation to the host and 0 signifies entirely non-optimal codon usage.

Let $|\alpha|$ represent the length of the protein $\alpha$, $f_i$ the frequency of the codon chosen for the $i$-th amino acid, and $\max(f_j)$ the maximum frequency over all synonymous codons for that amino acid. Codon frequencies are typically calculated using a reference mRNA transcript data for a particular organism. It is convenient to represent CAI in logarithmic form:

\begin{equation}\label{eq:log-cai}
    \log(\textsc{CAI}) = \frac{1}{|\alpha|}\sum_{i}{\log(\frac{f_i}{\max(f_j)})}
\end{equation}

To find the optimal CDS that balances MFE and codon usage, newer mRNA folding algorithms further extend RNA folding to incorporate CAI (\cite{zhang2023algorithm}, \cite{gu2024derna}). We adopt a similar notation to LinearDesign \cite{zhang2023algorithm} and combine $\log(\text{CAI})$ and MFE into a single objective score:

\begin{equation}\label{eq:cai-mfe}
    \textsc{CAIMFE} = \textsc{MFE}-\lambda\log(\textsc{CAI})
\end{equation}

Note that it is convenient to drop the $\frac{1}{|\alpha|}$ term from \Cref{eq:log-cai} when computing \textsc{CAIMFE}. Since \textsc{MFE} grows linearly with sequence length, it is natural to scale CAI by $\lambda\times |\alpha|$. Observe that $|\alpha| \times \frac{1}{|\alpha|}$ cancels.

Now we can define a combined MFE and CAI mRNA folding problem:  

\begin{equation}\label{eq:mfe-cai-fold}
    \textsc{Fold}(\alpha) = \operatorname{argmin}_{\pi \in \textsc{CDS}(\alpha)} \textsc{CAIMFE}(\pi, \alpha)
\end{equation}

\subsection{RNA Folding with Dynamic Programming}

RNA folding algorithms are based on the dynamic programming recursions of Zuker \& Stiegler \cite{zuker1981optimal}, while mRNA folding algorithms adapt these recursions with additional constraints. We begin by briefly describing the Zuker-Stiegler recursions and then outline the modifications applied in mRNA folding methods.

The energy functions used in RNA and mRNA folding algorithms are typically based on the nearest neighbor model \cite{turner2010nndb, mathews1999expanded, mathews2004incorporating, mittal2024nndb}. This thermodynamics-based model, derived from extensive optical melting experiments, has been in use since the 1970s \cite{studnicka1978computer} and is still under active development \cite{zuber2022nearest, mittal2024nndb}. All the mRNA folding algorithms described in this review utilize the nearest neighbor model (NN model) for their energy calculations.

The Zuker-Stiegler recursions (and the NN model) break up the energy calculation for an RNA structure into three kinds of ``loop'', based on the nearest neighbor model: loops that are closed by a single base pair, termed ``one loops''; loops that are closed by two base pairs, termed ``two loops''; and loops closed by more than 2 pairs, termed ``multiloops'' (\Cref{fig:fig1} panel C) \cite{mathews1999expanded, mathews2004incorporating,turner2010nndb, mittal2024nndb}.

We denote the energy contribution of a one loop by $\textsc{OneLoop}(i,j)$, where $(i,j)$ is the closing pair. Similarly, $\textsc{TwoLoop}(i,j,k,l)$ denotes the energy contribution of a two loop with $(i,j)$ and $(k,l)$ as the closing pairs. It is assumed that $i<k<l<j$. Multiloops are treated differently. The energy contribution of a multiloop is given by $\textsc{ML}_{\text{init}} + \textsc{ML}_{u} \times u + \textsc{ML}_{p} \times p$ where $u$ is the number of unpaired nucleotides enclosed by the loop, and $p$ is the number of pairs closing the loop. So, $\textsc{ML}_{\text{init}}$ is an initiation constant, $\textsc{ML}_{u}$ is the cost of an unpaired nucleotide, and $\textsc{ML}_{p}$ is the cost of a closing pair for the loop. See \cite{ward2017advanced} for a history of and justification for this multiloop model.

We have omitted several details of the modern NN model as they add complexity and obscure core ideas. These include helix end penalties, coaxial stacking, dangling ends, and terminal mismatches. Once the core ideas are understood, we think their addition should not be difficult. However, the reader should be aware that we do not cover them. We refer the reader to the Nearest Neighbor Database for a full description \cite{mittal2024nndb, turner2010nndb}.

\begin{figure}[h!]
    \centering
    \includegraphics[width=0.7\textwidth]{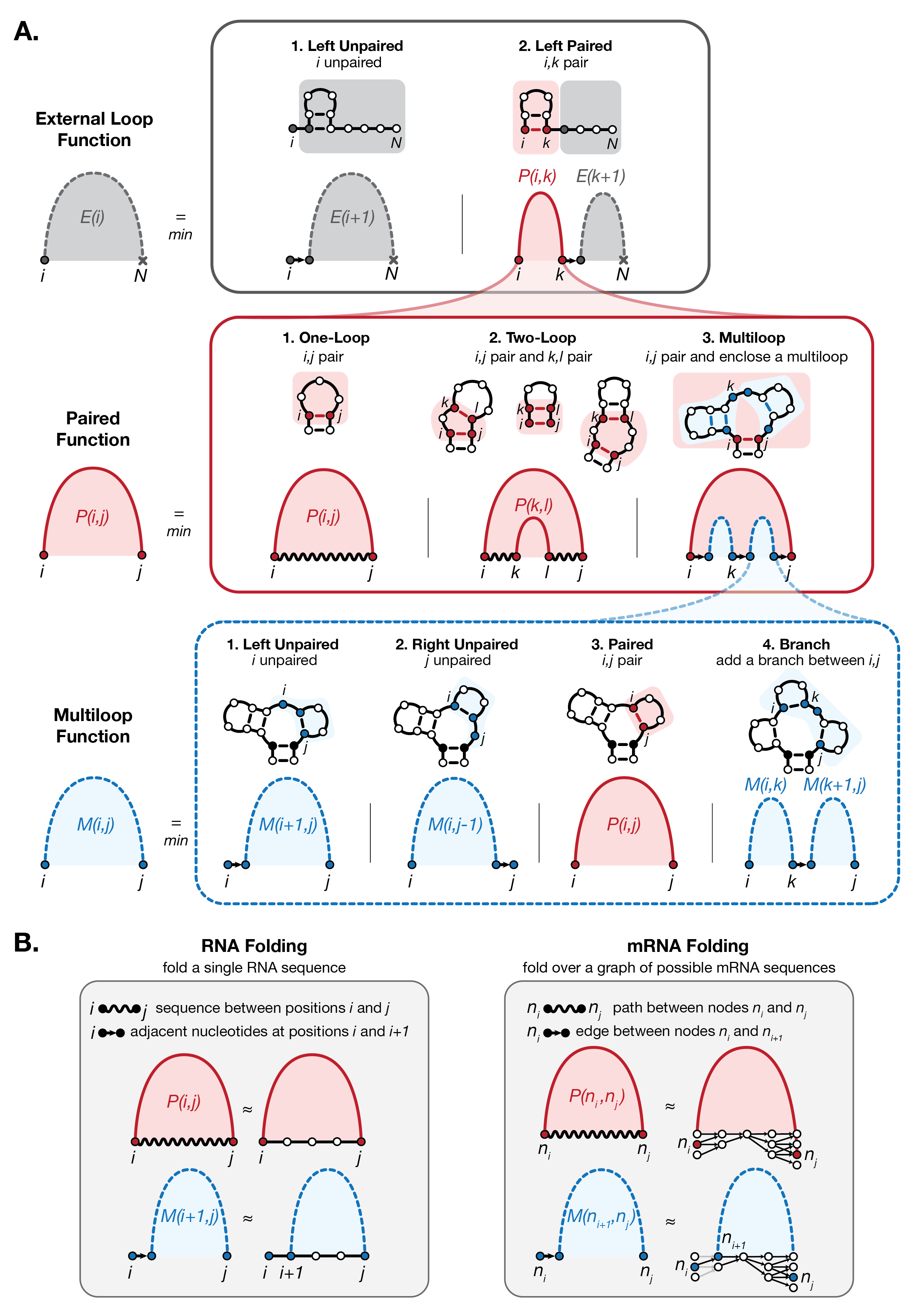}
    \caption{RNA Folding Recursions. (A) depicts the traditional Zuker-Stiegler RNA folding recursions. Each case of the recursions is represented by a Feynman-like diagram. The external loop function is represented by a dashed gray arc, the paired function is represented by a solid red arc, and the multiloop function is represented by a dashed blue arc. Solid black line segments between non-adjacent sequence indices represents a span of unpaired nucleotides. (B) depicts the codon folding recursions, which extend the Zuker-Stiegler recursions. Sequence indices are replaced with nodes at those indices. Straight black line segments are replaced with either wavy black line segments or black arrows. Wavy line segments represent a path between two nodes in the codon graph, while arrows represent an edge between two adjacent nodes in the graph.}
    \label{fig:fig2}
\end{figure}

The following dynamic programming recursions compute the MFE based on Zuker \& Stiegler's approach. \Cref{fig:fig2} panel A provides a graphical version of these.

\begin{equation}\label{eq:zuker-P}
    P(i,j) = \min \left\{
    \begin{array}{ll}
     \textsc{OneLoop}(i,j) \\
    \min_{k,l : i<k<l<j}{\textsc{TwoLoop}(i,j,k,l) + P(k,l)} \\
    \min_{k : i<k<j}{M(i+1,k) + M(k+1, j-1) + \textsc{ML}_{\text{init}} + \textsc{ML}_{p}} \\
    \end{array}
    \right.
\end{equation}

The paired function, $P(i,j)$, is the MFE over all substructures between $i$ and $j$ given that $i$ and $j$ are assumed to be paired. $P(i,j)=\infty$ if the nucleotides at $(i,j)$ cannot form a valid pair or if $i \geq j$.

\begin{equation}\label{eq:zuker-M}
    M(i,j) = \min \left\{
    \begin{array}{ll}
     M(i+1,j) + \textsc{ML}_{u} \\
     M(i,j-1) + \textsc{ML}_{u} \\
     P(i,j) + \textsc{ML}_{p} \\
     \min_{k : i<k<j}{M(i,k) + M(k+1, j)}\\
    \end{array}
    \right.
\end{equation}

The multiloop function, $M(i,j)$ is the MFE over all substructures between $i$ and $j$ given that there is at least one base pair in the substructure. Note that the pair does not necessarily need to be $(i,j)$. $M(i,j)=\infty$ when there could not be any pair (i.e., $i>j$).

\begin{equation}\label{eq:zuker-E}
    E(i) = \min \left\{
    \begin{array}{ll}
     E(i+1) \\
     \min_{k : i<k< N}{P(i,k) + E(k+1)}\\
    \end{array}
    \right.
\end{equation}

The external loop function, $E(i)$ is the MFE over all substructures for the suffix of nucleotides from $i$ to $N$ (where $N$ is the RNA length). The nucleotide $i$ is assumed to be in the external loop, which is the region not contained inside any base pair. Note that the external loop does not have an associated energy function in the nearest neighbor model, unlike one loops, two loops, and multiloops. The base case is $E(i)=0$ when $i>N$ where $N$ is the sequence length.

The $E$ function is used to extract the solution to the RNA folding problem, i.e., the MFE value. $E(1)$ is the MFE over all possible structures, assuming nucleotides are indexed from 1 to $N$.

\section{mRNA Folding with Dynamic Programming}\label{sec:mrna_folding}

mRNA folding algorithms in the literature can be divided into two types based on their approach to incorporating codon constraints into the folding process. The first type, introduced by Cohen and Skiena \cite{cohen2003natural} and later used by DERNA \cite{gu2024derna}, can be described as ``codon-constrained'' dynamic programming. The second type, introduced by CDSfold \cite{terai2016cdsfold} and refined by LinearDesign \cite{zhang2023algorithm},  we call ``codon graph'' dynamic programming.

Codon-constrained methods add codon constraints to the Zuker-Stiegler recursions. For example, $P(i,j)$ becomes $P(C_i, C_j, i, j)$. The semantics are similar, but incorporate assumptions about the codons that the $i$-th and $j$-th nucleotides are in. First, let $\textsc{Codon}(i)=\lfloor i/3 \rfloor$ represent the amino acid index that the $i$-th nucleotide corresponds to. Now, define $P(C_i, C_j, i, j)$ as the MFE over all substructures between $i$ and $j$ given that $i$ and $j$ are paired \textit{and where the codon at $\textsc{Codon}(i)$ is $C_i$ and the codon at $\textsc{Codon}(j)$ is $C_j$}. The other dynamic programming functions are generalized similarly.

``Codon graph'' methods use pointers into a graph instead of sequence indexes, enabling substantially more efficient mRNA folding algorithms. As these methods offer significant improvements over earlier codon-constrained approaches in terms of efficiency, simplicity, and extensibility, our focus will primarily be on them.

\subsection{Codon Graph Algorithms}

The first codon graph mRNA folding method was CDSfold \cite{terai2016cdsfold}. Conceptually, CDSfold works by computing the same tables in the Zuker-Stiegler algorithm, but $i$ and $j$ denote pointers into a graph instead of indexes into a sequence. This is an important conceptual shift and is the main idea that enables more efficient mRNA folding algorithms.

It should be stated that CDSfold does not explicitly use a codon graph \cite{terai2016cdsfold}. Instead, a nucleotide constrained version of the recursions (similar to codon constrained described above) is used. Then, ``extended nucleotides'' are introduced to deal with non-adjacent dependencies between nucleotides inside a codon. As Terai, Kamegai, and Asai point out, this is conceptually a graph \cite{terai2016cdsfold}. A contribution of this work is to formalize this notion by introducing the codon graph as an elegant way to describe the CDSfold and LinearDesign algorithms and unify them using the same underlying algorithmic framework.

\begin{figure}[h]
    \centering
    \includegraphics[width=1.0\textwidth]{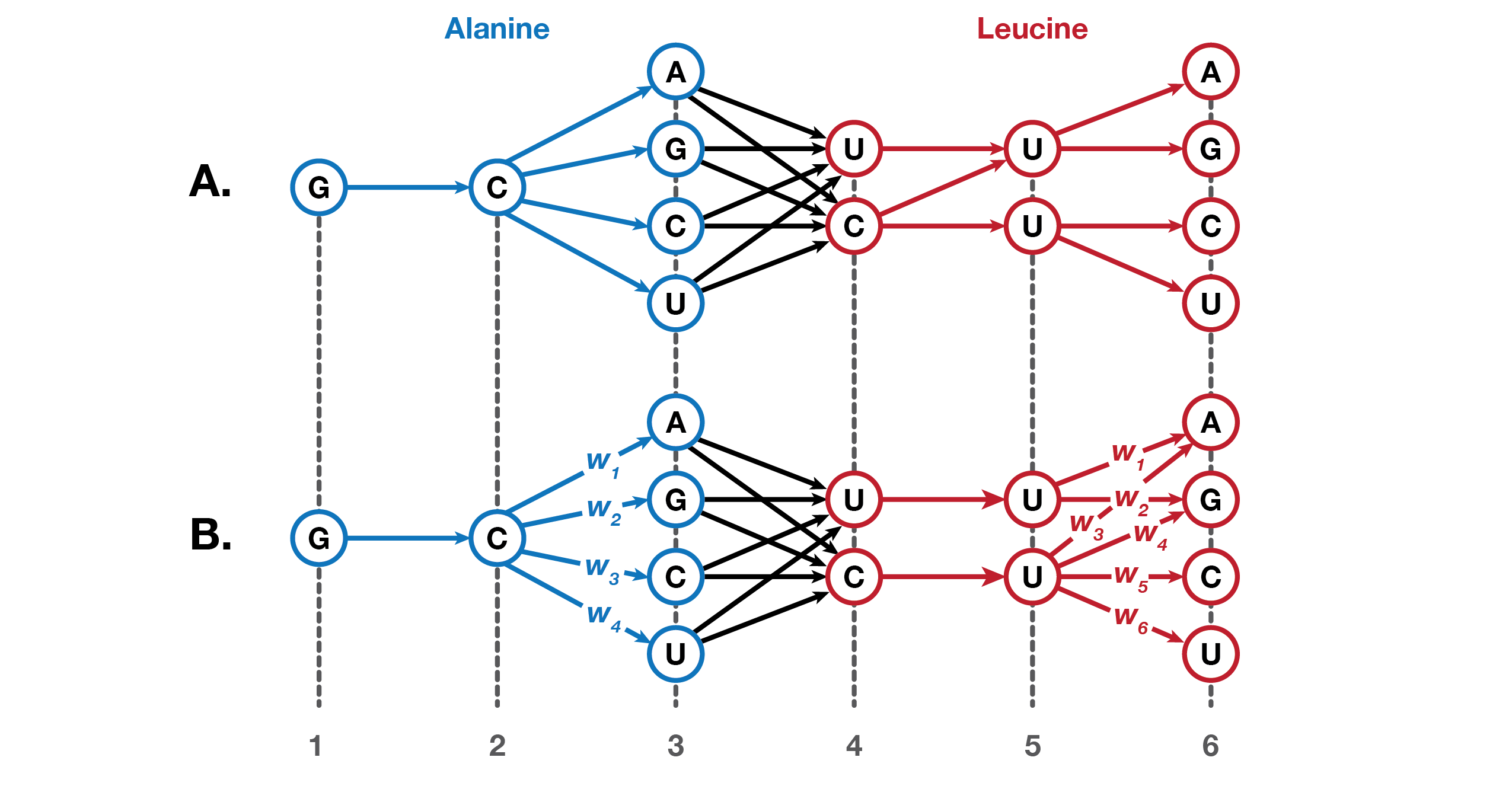}
    \caption{Codon Graphs. (A) depicts an ``extended nucleotide'' codon graph, as used by CDSfold. The codon subgraph for Alanine is on the left (blue, columns 1 to 3) with Leucine on the right (red, columns 4 to 6). A nucleotide (\texttt{A, U, G, C}) is associated with each vertex. The set of blue paths from left to right corresponds to the set of valid codons for Alanine, while the red paths correspond to the valid codons for Leucine. The two subgraphs are concatenated by the black edges. (B) depicts a modified codon graph with edge weights, as used by LinearDesign. A weight $w_i$ is associated with each of the rightmost edges in the Alanine and Leucine subgraphs, which corresponds to the weight of the corresponding codon. Since there is only a single codon path passing through each weighted edge, the corresponding codon is unambiguous.}
    \label{fig:fig3}
\end{figure}

In standard RNA folding, the sequence is fixed, so indexes into the sequence are sufficient to know which base identities are involved. This is important since the energy functions (e.g., $\textsc{OneLoop}(i,j)$ and $\textsc{TwoLoop}(i,j,k,l)$) depend on the base identities involved. In contrast, for mRNA folding, there is no fixed sequence. Instead, we are folding over all valid sequences. The solution employed by CDSfold is to construct a graph such that there is a one-to-one mapping between valid sequences and paths in the graph. Then, instead of an index, a pointer to a node in the graph can be used (see \Cref{fig:fig3}).

The CDSfold graph is constructed from ``extended nucleotide'' subgraphs for each amino acid in the protein. The ``extended nucleotide'' terminology refers to two nodes encoding the same nucleotide identity at the same sequence position, which captures dependencies between the first and last nucleotide in a codon. This only occurs for some codons, such as Serine, Arginine, and Leucine in the standard codon table.

The amino acid subgraph is constructed so that each path corresponds to a valid codon. Consider Leucine, which has six valid codons: \texttt{CUC, CUU, CUA, CUG, UUA, UUG}. We can construct each of these six codons by following a different path through the Leucine subgraph (see the red subgraph in \Cref{fig:fig3} panel A). We can then construct the protein graph by concatenating individual amino acid subgraphs. \Cref{fig:fig3} panel A shows how the CDSfold construction concatenates the Alanine and Leucine subgraphs by adding every possible edge from the end of Alanine to the start of Leucine as depicted by the black edges.

 We refer to these graph constructions (in \Cref{fig:fig3}) as \textit{codon graphs}.
 
 First, we need some definitions for accessing the codon graph. Let $u$ refer to a node in the codon graph. Define $\operatorname{out}(u)$ as the set containing $u$'s neighbours---this corresponds to the outgoing edges from $u$ in the codon graph. Similarly, let $\operatorname{in}(u)$ be the neighbours in the codon graph where edge directions are reversed---this corresponds to the incoming edges to $u$. Let $\operatorname{atpos}(i)$ be the set of nodes that correspond to the $i$-th sequence position. Note that a codon graph can contain several nodes at the same position in an mRNA. In \Cref{fig:fig3}, $\operatorname{atpos}(i)$ corresponds to the set of nodes at the $i$-th column. Let $b_u$ denote the base identity (A, U, G, or C) associated with node $u$.

 Observe that in \Cref{fig:fig3} some nodes cannot be reached from other nodes. Consider two nodes in the codon graph $u$ and $v$. We say that $u$ can reach $v$ if there is a directed path in the codon graph from $u$ to $v$. We can construct a reachability table $R(u,v)$ which is true if $u$ can reach $v$ and false otherwise. $R$ can be constructed efficiently using standard graph algorithms: e.g., a depth-first search from each node.

 Consider the three cases of \Cref{eq:zuker-P}. They can be modified to operate on the codon graph as follows.

\begin{equation}\label{eq:cdsfold-P} 
    P(n_i,n_j) = \min \left\{
    \begin{array}{ll}
    & \textsc{OneLoop}(b_{n_i},b_{n_j}),
    \\\\
    
    \min\limits_{\substack{k,l : i<k<l<j, \\ n_{k} \in \text{atpos}(k) : R(n_i, n_k), \\ n_{l} \in \text{atpos}(l) : R(n_l, n_j)}}
    & \textsc{TwoLoop}(b_{n_i},b_{n_j},b_{n_{k}},b_{n_{l}}) + P(n_k, n_l),
    \\\\
    
    \min\limits_{
    \substack{k : i<k<j,
    \\ n_k \in \text{atpos}(k),
    \\ n_{k+1} \in \text{out}(n_k)
    \\ n_{i+1} \in \text{out}(n_i),
    \\ n_{j-1} \in \text{in}(n_j)
    }
    }
    & M(n_{i+1},n_k) + M(n_{k+1}, n_{j-1}) +\textsc{ML}_{\text{init}} + \textsc{ML}_{\text{p}}
    \end{array}
    \right.
\end{equation}

\Cref{eq:cdsfold-P} operates on graph nodes instead of sequence indexes. In particular, $i$ and $j$ are replaced with $n_i$ and $n_j$, which represent a codon graph node at RNA indexes $i$ and $j$ respectively. In each case, we must try all possible nodes that could be at the sequence indexes in the former recurrence, \Cref{eq:zuker-P}. Note that $R(n_i,n_j)$, $\operatorname{out}(n_i)$, $\operatorname{in}(n_j)$, and similar are used to ensure that these nodes are reachable in the codon graph. The recursions for $M$ and $E$ follow similar patterns.

\begin{equation}\label{eq:cdsfold-M} 
    M(n_i,n_j) = \min \left\{
    \begin{array}{ll}
    \min\limits_{\substack{n_{i+1} \in \text{out}(n_i)}}
    & M(n_{i+1}, v) + \textsc{ML}_{u},
    \\\\
   \min\limits_{\substack{n_{j-1} \in \text{in}(n_j)}}
    & M(n_i, n_{j-1}) + \textsc{ML}_{u},
    \\\\
    & P(n_i,n_j) + \textsc{ML}_{p},
    \\\\
    \min\limits_{\substack{k : i<k<j, \\ n_k \in \text{atpos}(k), \\ n_{k+1} \in \text{out}(n_k)}}
    & M(n_i,n_k) + M(n_{k+1}, n_j)
    \end{array}
    \right.
\end{equation}

\begin{equation}\label{eq:cdsfold-E} 
    E(n_i) = \min \left\{
    \begin{array}{ll}
    \min\limits_{\substack{n_{i+1} \in \text{out}(n_i)}}
    & E(n_{i+1}),
    \\\\
    \min\limits_{\substack{k : i<k<N, \\ n_k \in \text{atpos}(k), \\ n_{k+1} \in \text{out}(n_k)}}
    & P(n_i,n_k) + E(n_{k+1})
    \end{array}
    \right.
\end{equation}

We use $N$ to denote the RNA length in \Cref{eq:cdsfold-E}. This is always $3 \times |\alpha|$ where $|\alpha|$ is the input protein length.

The base cases for these recursions are similar to the Zuker-Stiegler versions. $P(n_i, n_j)=\infty$ when the nucleotides corresponding to nodes $n_i$ and $n_j$ cannot form a valid base pair or if there is no path from $n_i$ to $n_j$: $\lnot R(n_i, n_i)$. Also, $M(n_i,n_j)=\infty$ if there is no path from $n_i$ to $n_j$. The base case for $E$ needs some extra work, since previously $E(i>N)=0$. We introduce a special ``end node'' $\omega$ to the codon graph at index $N+1$. There are edges from all nodes at $N$ to $\omega$. We define $E(\omega)=0$.

\Cref{eq:cdsfold-P}, \Cref{eq:cdsfold-M}, and \Cref{eq:cdsfold-E} are equivalent to the CDSfold recursions \cite{terai2016cdsfold}. See \Cref{fig:figS1} in the Supplementary Material for a visual description of the recursions. Our presentation simplifies them by introducing the concept of an explicit codon graph, but the underlying ideas are the same. Next, we will extend our dynamic programming on a codon graph framework to explain how LinearDesign operates \cite{zhang2023algorithm}.

\subsection{Incorporating CAI} \label{sec:incorporating-cai}

LinearDesign improved on CDSfold to enable simultaneous optimization of both CAI and MFE as in \Cref{eq:cai-mfe} \cite{zhang2023algorithm}. The LinearDesign algorithm is described in terms of a deterministic finite-automata and context-free grammar parsing. These ideas correspond to the codon graph and dynamic programming in our framework. While different nomenclature is used, the resulting algorithms are equivalent. Our codon graph and dynamic programming framework helps us to put LinearDesign into context with existing algorithms such as the Zuker-Stiegler algorithm \cite{zuker1981optimal} and CDSfold \cite{terai2016cdsfold}.

A complete description of the LinearDesign algorithm does not appear in the literature, as \cite{zhang2023algorithm} provides only a description of the algorithm on a simplified model. Specifically, the Nussinov model \cite{nussinov1980fast} is used, which is much simpler than the full NN model. A major contribution of this work is to provide a full description of the algorithm. LinearDesign uses a beam search heuristic adapted from LinearFold \cite{huang2019linearfold} to speed up execution at the cost of approximating the solution. We do not include this, as our goal is to provide the foundational mRNA folding algorithms without added heuristics.

LinearDesign incorporated CAI by modifying the codon graph with added edge weights. The graph for each amino acid is modified so that the path for each codon has at least one unique edge (\Cref{fig:fig3} panel B). This modifies the amino acid graphs from CDSfold. For example, compare the Leucine subgraphs in panel A of \Cref{fig:fig3} to that in panel B. In the LinearDesign construction, there is a unique edge for each of the 6 codons between the middle (U, U) and rightmost (C, U, A, G) columns. In general, it is possible to construct the LinearDesign amino acid graphs by constructing a path for each unique codon prefix (e.g., \texttt{CU} and \texttt{UU} for Leucine), then adding edges for all final nucleotides in each codon. Note that we use the standard codon table in our discussion, but in theory this method extends to arbitrary codon tables.

By construction, each rightmost edge in the LinearDesign amino acid graph corresponds to a single codon. CAI is incorporated into the graph by adding weights to these edges equal to the contribution of the corresponding codon to the total weighted log-CAI: $-\log(\frac{f_i}{\max(f_i)}) \times \lambda$ from \Cref{eq:cai-mfe}. Note that other edges are not assigned a log-CAI weight and are assumed to have a weight of zero. Each path corresponds to a valid CDS and the sum of weights on the path corresponds to $-\log(\textsc{CAI}) \times \lambda$ for that CDS.

A significant difference from the prior recursions is that a path between nodes can contribute a weight even if there are no paired nucleotides involved. For example, the $\textsc{TwoLoop}$ case in \Cref{eq:cdsfold-P} only checked $R(n_i,n_k)$ for the stretch of unpaired nucleotides from $n_i$ to $n_k$. However, since some of the edges in a path from $n_i$ to $n_k$ could be weighted, we must now incorporate the weight.

Define $\text{LCAI}(u \leadsto v)$ as the sum of log-CAI weights on a minimum-weight path from node $u$ to node $v$.  Let $\text{LCAI}(u \leadsto v)=\infty$ if there is no path. All values for $\text{LCAI}(u \leadsto v)$ can be pre-computed and stored in a table. There are several ways to do this including dynamic programming on the graph (since it is directed and acyclic), or using standard shortest path algorithms on the graph. Johnson's algorithm can compute the all-pairs shortest paths with negative edge weights \cite{johnson1977efficient}. All such methods are asymptotically dominated by the cost of the remainder of the algorithm.

\begin{equation}\label{eq:lineardesign-P} 
    P(n_i,n_j) = \min \left\{
    \begin{array}{ll}
    & \textsc{OneLoop}(b_{n_i},b_{n_j}) + \text{LCAI}(n_i \leadsto n_j),
    \\
    
    \min\limits_{
    \substack{k,l : i<k<l<j,
    \\ n_{k} \in \text{atpos}(k),
    \\ n_{l} \in \text{atpos}(l)
    }
    }
    &
    \begin{aligned}
    \\
    & \textsc{TwoLoop}(b_{n_i},b_{n_j},b_{n_{k}},b_{n_{l}}) + P(n_k, n_l) \\
    & + \text{LCAI}(n_i \leadsto n_k) + \text{LCAI}(n_l \leadsto n_j),
    \end{aligned}
    \\
    
    \min\limits_{
    \substack{
    k : i<k<j,
    \\ n_k \in \text{atpos}_{k},
    \\ n_{k+1} \in \text{out}_{n_k},
    \\ n_{i+1} \in \text{out}(n_i),
    \\ n_{j-1} \in \text{in}(n_j)
    }
    } &
    \begin{aligned}
    \\
    & M(n_{i+1},n_k) + M(n_{k+1}, n_{j-1}) +\textsc{ML}_{\text{init}} + \textsc{ML}_{\text{p}} \\
    & + \textsc{LCAI}(n_i \leadsto n_{i+1}) + \textsc{LCAI}(n_k \leadsto n_{k+1}) + \textsc{LCAI}(n_{j-1} \leadsto n_j)
    \end{aligned}
    \end{array}
    \right.
\end{equation}

\Cref{eq:lineardesign-P} updates $P$ from \Cref{eq:cdsfold-P} to incorporate edge weights. The update rule is that $\text{LCAI}(u \leadsto v)$ is added when the recursions consider a transition between codon graph nodes that will not be considered in a recursive call. Note that $\text{LCAI}(u \leadsto v)$ is used even when the path is only a single edge, e.g., $\textsc{LCAI}(n_k \leadsto n_{k+1})$ in \Cref{eq:lineardesign-P}.

\begin{equation}\label{eq:lineardesign-M} 
    M(n_i,n_j) = \min \left\{
    \begin{array}{ll}
    \min\limits_{\substack{n_{i+1} \in \text{out}(n_i)}}
    & M(n_{i+1}, n_j) + \text{ML}_{u} + \text{LCAI}(n_i \leadsto n_{i+1}),
    \\\\
   \min\limits_{\substack{n_{j-1} \in \text{in}(n_j)}}
    & M(n_{j-1}, n_j) + \text{ML}_{u} + \text{LCAI}(n_{j-1} \leadsto n_j),
    \\\\
    & P(u,v) + \text{ML}_{p},
    \\\\
    \min\limits_{\substack{k : i<k<j, \\ n_k \in \text{atpos}(k), \\ n_{k+1} \in \text{out}(n_k)}}
    & M(n_i,n_k) + M(n_{k+1}, n_j) + \text{LCAI}(n_{k} \leadsto n_{k+1})
    \\\\
    \end{array}
    \right.
\end{equation}

\begin{equation}\label{eq:lineardesign-E} 
    E(n_i) = \min \left\{
    \begin{array}{ll}
    \min\limits_{\substack{n_{i+1} \in \text{out}(n_i)}}
    & E(n_{i+1}) + \text{LCAI}(n_i \leadsto n_{i+1}),
    \\\\
    \min\limits_{\substack{k : i<k<n, \\ n_k \in \text{atpos}(k), \\ n_{k+1} \in \text{out}(n_k)}}
    & P(n_i,n_k) + E(n_{k+1}) + \text{LCAI}(n_{k} \leadsto n_{k+1})
    \end{array}
    \right.
\end{equation}

The recursions for $M$ and $E$ are similarly updated in \Cref{eq:lineardesign-M} and \Cref{eq:lineardesign-E}. The base cases for all recursions are unchanged.

\subsection{Traceback}

The recursions presented compute the score of the optimal solution but do not construct the solution itself. As is typical for dynamic programming algorithms, the solution can be recovered using a \textit{traceback}. The traceback is a standard procedure that recovers the solution by recapitulating the steps in the recursions that led to the best score \cite{eddy2004dynamic}. In the case of the algorithms presented here, the goal is to recover the mRNA. The traceback details are tedious and mechanistic, but are provided in the Supplementary Material (see \Cref{alg:traceback}) for completeness.

\subsection{Additional Energy Model Details}
Some details were omitted from the prior description of our dynamic programming algorithm for brevity. We assumed that the $\textsc{TwoLoop}(b_{n_i}, b_{n_j}, b_{n_k}, b_{n_l})$ energy function only needs to know the base identities of the two closing base pairs $(i,j)$ and $(k,l)$. However, for the full energy model, this is not true. In general, the energy function may need to know the mismatched base's identities at positions $(i+1,j-1)$ and $(k-1, l+1)$. The full form of the energy function is $\textsc{TwoLoop}(b_{n_i}, b_{n_j}, b_{n_{i+1}}, b_{n_{j-1}}, b_{n_k}, b_{n_l}, b_{n_{k-1}}, b_{n_{l+1}})$.

This does not change the dynamic programming recursion's structure, but it does complicate them. In particular, we modify $P(n_i,n_j)$ by taking the minimum over $n_{i+1} \in \operatorname{out}(n_i)$, $n_{j-1} \in \operatorname{in}(n_j)$, $n_{k-1} \in \operatorname{in}(n_k)$, and $n_{l+1} \in \operatorname{out}(n_l)$. There are special cases when $i+1=k-1$, $i=k-1$, $j-1=l+1$, or $j=l+1$. These conveniently correspond to specific special cases in the NN model including ``stacks'', ``bulges'', and ``1xn'' internal loops.

We also assumed that ``hairpin loops'' can be described by the simple function $\textsc{OneLoop}(b_{n_i}, b_{n_j})$. The full energy model takes the mismatch into account, so we must use $\textsc{OneLoop}(b_{n_i}, b_{n_j}, b_{n_{i+1}}, b_{n_{j-1}})$. This requires a similar modification as for two loops. In addition, there are ``special'' hairpin loops, which are specific sequences that have a unique energy term. Since these are small (3, 4, and 6 unpaired nucleotides in length), they can be incorporated into the algorithm by brute-force enumeration. That is, when computing $P(n_i,n_j)$, if $i-j-1 \leq 6$, we enumerate all paths from $n_i$ to $n_j$. Each of these paths is a possible hairpin sequence, and we take the minimum over all sequences. Sequences corresponding to special hairpins use the special hairpin rule, otherwise $\textsc{OneLoop}(b_{n_i}, b_{n_j}, b_{n_{i+1}}, b_{n_{j-1}})$ is used.

The reader is referred to the Nearest Neighbor Database for more details on hairpin loops, internal loops, stacks, and bulges in the energy model \cite{turner2010nndb, mittal2024nndb}.

\subsection{Complexity Analysis}

In calculating the computational complexity of the algorithms we assume that tables are used to store solutions for $P$, $M$, and $E$ and these are filled bottom-up \cite{eddy2004dynamic}. This is similar to the implementation of existing mRNA folding algorithms \cite{terai2016cdsfold, zhang2023algorithm}. For completeness, a valid bottom-up fill order is to iterate backwards through 5' sequence indexes $i$ and for each $i$ iterate forward through 3' sequence indexes $j$. 

Let $N=3\times|\alpha|$ be the mRNA length. The table size for $P$ is $O(N^2)$. Each mRNA sequence index has at most four nodes (one for each nucleotide) using the construction from \Cref{sec:incorporating-cai} assuming the standard codon table. So, the codon graph contains an upper bound of $N\times 4$ nodes. The total number of table entries is bounded by $O(N^2)$ combinations of the nodes $n_i$ and $n_j$. The cost of computing the solution for a table entry is dominated by iterating through all $O(N^2)$ combinations of $n_k$ and $n_l$. However, in RNA folding algorithms it is typical to limit the size of two loops to at most 30 unpaired nucleotides, as they rapidly become thermodynamically unfavourable \cite{lyngso1999internal}. In this case, the calculation is dominated by the cost of considering multiloop splits, which involves enumerating all pairs of nodes $n_k$ and $n_{k+1}$. There are at most $O(N)$ such pairs, since there are at most four options for $n_{k+1}$. This gives a time complexity of $O(N^3)$.

The table for $M$ is similarly $O(N^2)$ in size. It is also dominated by calculating splitting pairs $n_k$ and $n_{k+1}$. As such, the total time complexity for filling $M$ is $O(N^3)$.

The table for $E$ is $O(N)$ in size since it is parameterized by a single node. The worst case cost of calculating an entry is $O(N)$, as it similarly considers all splitting pairs $n_k$ and $n_{k+1}$. As such, the total time complexity for filling $E$ is $O(N^2)$.

The algorithm is dominated by filling the $M$ and $P$ tables. The worst case time complexity is $O(N^3)$ and the space complexity is $O(N^2)$. The cost of computing the shortest path table $\textsc{LCAI}(u \leadsto v)$ using an efficient algorithm such as Johnson's algorithm \cite{johnson1977efficient} is at most $O(N^2 \log N)$, since an upper bound on the number of nodes in the graph is $N\times 4$ and an upper bound on the number of edges is $N \times 4^2$. The traceback is similarly dominated, since it will only visit table entries in the optimal solution and its total time cost must be less than the cost of computing the tables.

\subsection{Pareto Optimality}

DERNA \cite{gu2024derna} introduced a unique feature to mRNA folding algorithms to find all Pareto optimal mRNAs. An mRNA $\pi$ is Pareto optimal if there is no sequence $\pi'$ that dominates $\pi$ in terms of both CAI and MFE: $\nexists_{\pi'} : \textsc{CAI}(\pi') > \textsc{CAI}(\pi) \land \textsc{MFE}(\pi') < \textsc{MFE}(\pi)$. In other words, a Pareto optimal set for mRNA folding solutions contains one mRNA for each achievable CAI value and that sequence must have the minimum MFE possible for that CAI.

Finding all Pareto optimal mRNA sequences solves a problem in LinearDesign \cite{zhang2023algorithm}. The term $\lambda$ is used to balance CAI and MFE in \Cref{eq:cai-mfe}. Selecting the right $\lambda$ can be challenging. For instance, if an mRNA designer wants to find a sequence with $\textsc{CAI}>0.9$, then they must run LinearDesign multiple times to binary search the lowest $\lambda$ that satisfies the condition. The set of Pareto optimal solutions contains a solution for every possible tradeoff between CAI and MFE.

The recursions presented in this work can be modified to compute all Pareto optimal solutions. Each of $P(u,v)$, $M(u, v)$, and $E(u)$ computes the CAIMFE (defined in \Cref{eq:cai-mfe}) of the optimal solution to the corresponding sub problem. Instead, they could compute a set of Pareto optimal solutions for each sub problem. For example, the dynamic programming table for $P(u,v)$ might store a list of all $(\textsc{CAI}, \textsc{MFE})$ pairs for Pareto optimal solutions. Two lists can be combined by enumerating all pairs of elements (one from each list) and taking only Pareto optimal combinations. This is a straightforward, albeit naive, solution.

DERNA uses a more sophisticated but less pedagogically clear weighted sum method that exploits the convexity of the CAI-MFE tradeoff---increasing CAI monotonically increases MFE. Both methods could be adapted to the recursions presented here. DERNA extends the less-efficient codon constrained method for mRNA folding. This makes DERNA significantly slower than both CDSfold and LinearDesign, even when not running in Pareto optimal mode.

\subsection{Untranslated Regions}\label{sec:utrs}

An mRNA designer usually considers three regions: the 5' untranslated region (UTR), the CDS, and the 3' UTR. However, existing mRNA folding algorithms optimise only the CDS. Zhang \textit{et al.} \cite{zhang2023algorithm} suggested that an MFE-optimised CDS is less likely to have base pairs that interact with the UTRs, which is important to avoid any disruption in UTR function. This is especially important for the 5' UTR, since structure near the mRNA 5' end can substantially impair translation initiation \cite{babendure2006control}. There is some experimental evidence for this \cite{leppek2022combinatorial}, but the algorithms do not guarantee it.

\subsection{Structural Constraints}

The goal of MFE optimization in mRNA folding is to increase stability, which generally increases structure. However, in some cases it is desirable to suppress structure. For instance, reduced structure in the 5' UTR, particularly near the start codon, is associated with increased expression \cite{hinnebusch2016translational, sample2019human, ringner2005folding}. As mentioned in \Cref{sec:utrs}, it may be useful to avoid base pairs between the CDS and the UTRs. Also, it can be useful to avoid long helices since they can trigger an innate immune response \cite{liu2008structural}. To these ends it would be useful to extend mRNA folding algorithms to incorporate structural constraints.

CDSfold included a heuristic to discourage base pairs in a user-specified region \cite{terai2016cdsfold}. This heuristic penalizes $P(n_i,n_j)$ whenever $(i, j)$ is a suppressed base pair, reducing but not entirely eliminating their occurrence. By modifying the free energy landscape, it increases the free energy of any structure with a suppressed pair. However, the mRNA folding algorithm may still find a sequence with low MFE in the changed free energy landscape that can be even lower when suppressed pairs are allowed again. To address this, CDSfold heuristic also employs a second phase inspired by Gaspar \textit{et al.} \cite{gaspar2013mrna}. While effective for suppressing base pairing in specific regions, this heuristic does not generalize beyond that function and is not implemented with CAI optimization.

LinearDesign also uses a heuristic to avoid structure around the 5' UTR \cite{zhang2023algorithm} by excluding the first three codons and optimizing the remaining CDS. Then, all combinations for the three excluded codons are enumerated and evaluated. This method appears to work for reducing structure around the start codon, but does not scale to large regions (due to brute force enumeration) or generalize to arbitrary structural constraint. LinearDesign also avoids long helices, but their avoidance heuristic were not specified.

\subsection{Sequence Constraints}

Avoidance of certain sequences in an mRNA can be important. Factors like restriction enzyme sequences, repeated subsequences, and the proportion of G and C nucleotides can affect mRNA efficacy and ease of manufacturing \cite{metkar2024tailor}.

Zhang \textit{et al.} \cite{zhang2023algorithm} note that LinearDesign's deterministic finite automaton (DFA) can be modified to avoid certain motifs such as restriction enzyme recognition sequence \texttt{GGUACC}. Since the DFA is equivalent to the codon graph framework, these modifications translate directly. While similar modifications could be made by hand for other excluded sequences, this may become cumbersome if multiple excluded sequences overlap.

\section{Comparison of Existing Software Packages}\label{sec:benchmarks}

We conducted a series of experiments to compare existing mRNA folding software packages including LinearDesign \cite{zhang2023algorithm}, CDSfold \cite{terai2016cdsfold}, and DERNA \cite{gu2024derna}. These were downloaded from their respective GitHub repositories using commits \texttt{f0126ca}, \texttt{06f3ee8}, and \texttt{ac84b6f} compiled from source on Ubuntu 24.04 using GCC 13.2.0. All experiments were performed on the same Ubuntu system equipped with an AMD 7950X processor. The \textit{Homo sapiens} codon frequency table from the Kazusa database was used for all experiments \cite{nakamura2000codon}. All of our benchmarking results and code is available at \url{https://github.com/maxhwardg/mrna_folding_comparison}.

An overview of our findings is summarized in \Cref{tab:software_comparison}.

\begin{table}[h!]
\centering
\begin{tabular}{|l|l|l|l|l|l|l|l|}
\hline
\textbf{Software Package} & \textbf{Speed} & \textbf{Memory Usage} &  \textbf{CAI} & \textbf{Bugs} & \textbf{Approximate} & \textbf{Pareto Optimal} \\ \hline
LinearDesign \cite{zhang2023algorithm} & \underline{Fast} & High & \underline{Yes} & Observed & Yes & No \\ \hline
DERNA \cite{gu2024derna} & Slow & High & \underline{Yes} & Observed & \underline{No} & \underline{Yes}  \\ \hline
CDSfold \cite{terai2016cdsfold} & Intermediate & \underline{Low} & No & \underline{Not observed} & \underline{No} & No \\ \hline
\end{tabular}
\caption{Comparison of mRNA folding software packages. Underlined entries are the most desirable quality for the corresponding column.}
\label{tab:software_comparison}
\end{table}

\subsection{Performance Benchmarks}

\begin{figure}[h]
    \centering
    \begin{subfigure}[b]{0.45\textwidth}
        \centering
        \includegraphics[width=\textwidth]{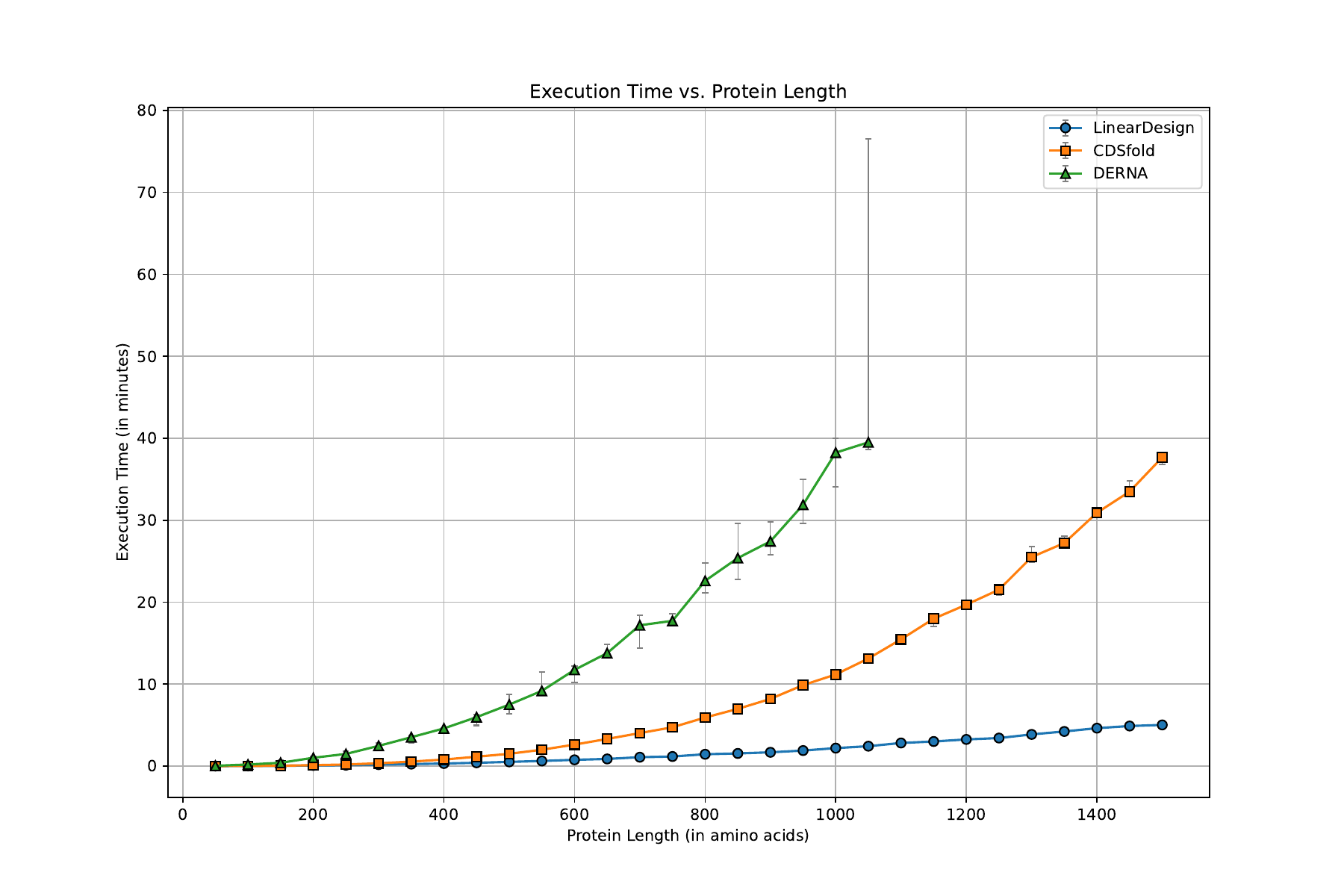} 
        \caption{Execution time}
        \label{fig:random_seq_bench_time}
    \end{subfigure}
    \hfill
    \begin{subfigure}[b]{0.45\textwidth}
        \centering
        \includegraphics[width=\textwidth]{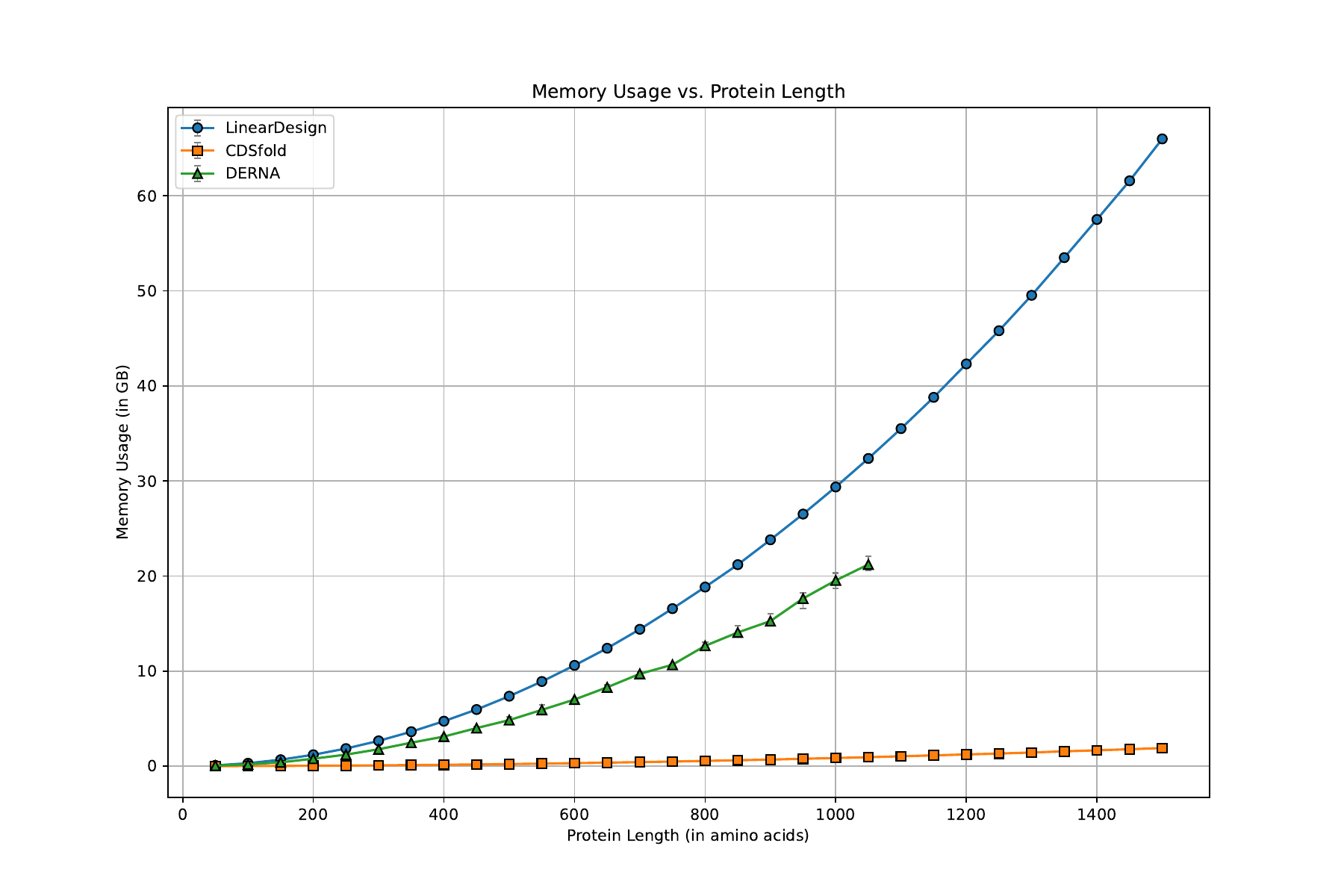} 
        \caption{Memory usage}
        \label{fig:random_seq_bench_mem}
    \end{subfigure}
    \caption{Benchmarks of available software packages for mRNA folding on randomly generated proteins. Execution time (a) and memory usage (b) of LinearDesign (blue circles), CDSfold (orange squares), and DERNA (green triangles) were measured for randomly generated protein sequences ranging from 50 to 1500 amino acids in length, with a stride of 50. Each data point represents the median execution time across three runs, with error bars representing the highest and lowest measure.}
    \label{fig:random_seq_bench}
\end{figure}

\begin{figure}[h]
    \centering
    \begin{subfigure}[b]{0.45\textwidth}
        \centering
        \includegraphics[width=\textwidth]{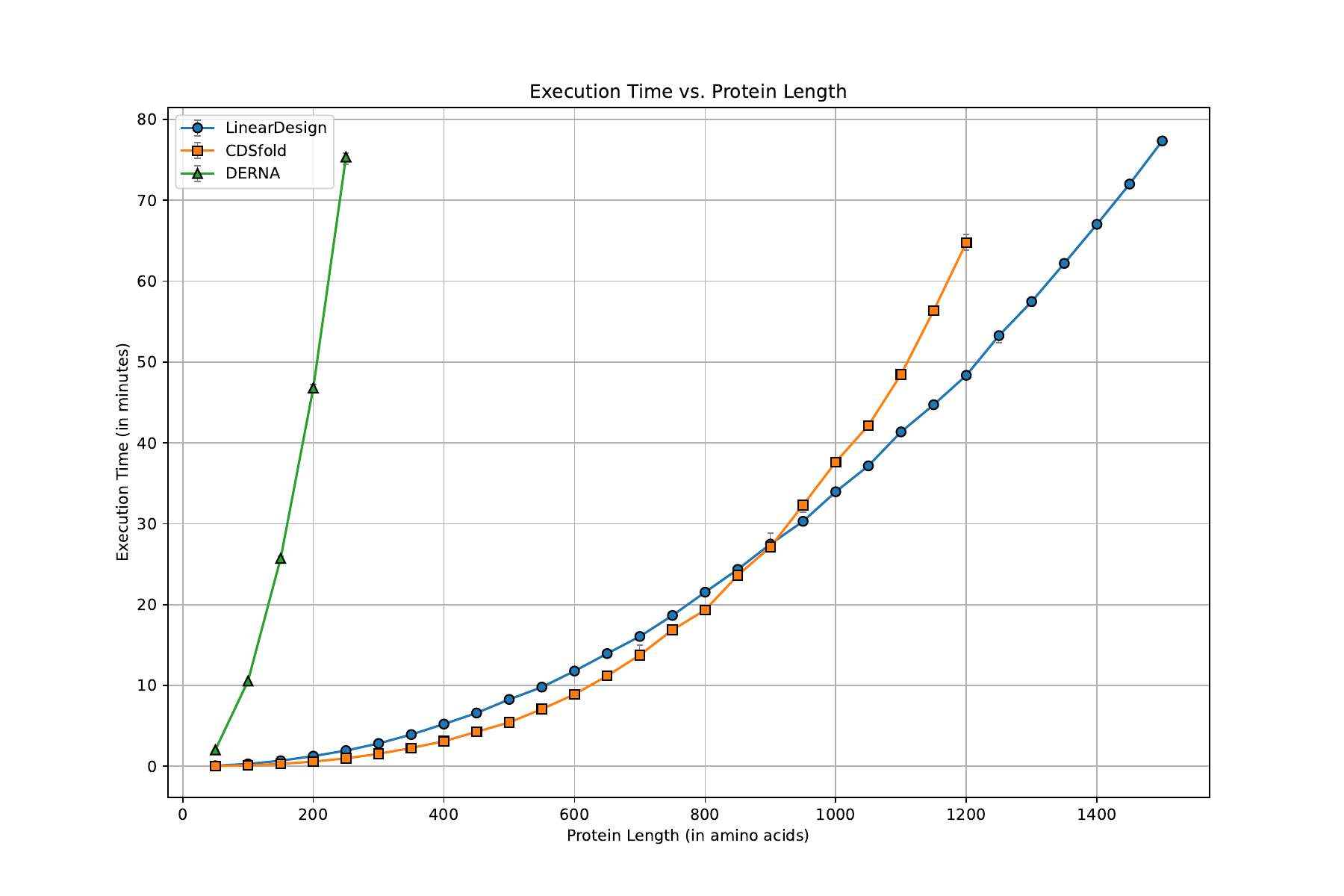} 
        \caption{Execution time}
        \label{fig:ml_bench_time}
    \end{subfigure}
    \hfill
    \begin{subfigure}[b]{0.45\textwidth}
        \centering
        \includegraphics[width=\textwidth]{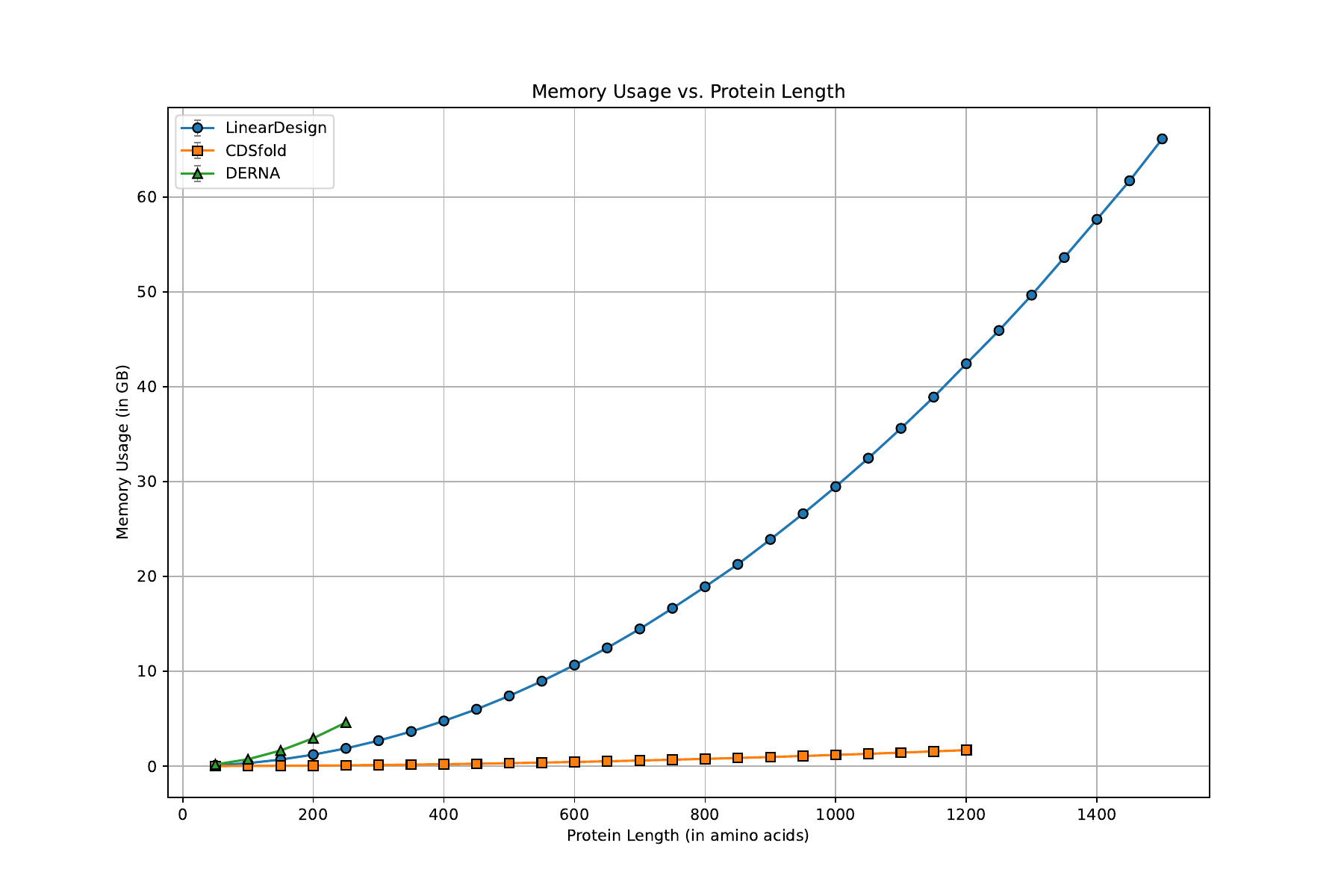}
        \caption{Memory usage}
        \label{fig:ml_bench_mem}
    \end{subfigure}
    \caption{Benchmarks of available software packages for mRNA folding on the amino acid sequence \texttt{MLLL\dots}} Execution time (a) and memory usage (b) of LinearDesign (blue circles), CDSfold (orange squares), and DERNA (green triangles) was measured. Since Leucine has the maximum synonymous codons (6), this provided a challenging scenario test set for these algorithms.
    \label{fig:ml_bench}
\end{figure}

The software packages were benchmarked on proteins ranging 50 and 1500 amino acids in length, with a stride of 50. Benchmarking of a software package was terminated if it exceeded a time of one hour. The protein sequences were randomly generated with uniformly sampled amino acids. Since CDSfold does not optimise CAI, LinearDesign was run with $\lambda=0$ and DERNA with $\lambda=1$ to emulate the behavior of CDSfold. The results are displayed in \Cref{fig:random_seq_bench}.

Benchmarking on random protein data (\Cref{fig:random_seq_bench}) shows that for execution time, LinearDesign was the fastest, followed by CDSfold, and DERNA. LinearDesign operates as an approximate algorithm, whereas CDSfold and DERNA are exact. During benchmarking, LinearDesign occasionally produced less optimized results (higher MFE) than the other algorithms, though such cases were rare and the differences were minor. An example sequence is available in our GitHub repository.

Memory usage followed a different trend: LinearDesign consumed the most memory, closely followed by DERNA, while CDSfold used minimal resources. Notably, LinearDesign required over 60GB of memory to fold a 1,450 amino acid protein.

The performance of mRNA folding algorithms is sensitive to protein composition. For instance, the sequence \texttt{MLLL\dots} (one methionine followed by a variable number of leucines) has a more challenging codon graph than a randomly created protein as leucine has the maximum number of synonymous codons (6). To evaluate this effect, we ran a second benchmark using various lengths of the \texttt{MLLL\dots} sequence. The results are displayed in \Cref{fig:ml_bench}. As expected, all software packages performed slower on this benchmark, with DERNA being particularly affected---it required 75 minutes to fold a 220 amino acid protein, compared to approximately one minute for both LinearDesign and CDSfold. CDSFold outperformed LinearDesign for lengths up to 900 amino acids for execution time, but LinearDesign was faster for longer sequences. The memory usage is higher with DERNA using relatively more memory.

\subsection{Software Bugs}

During the benchmarking process several bugs were found in the software packages. Most notably, the CAI values reported by DERNA did not match those produced by LinearDesign and by our CAI calculator, despite all programs using the same codon frequency table. Additionally, DERNA showed non-deterministic behavior, and reported different CAI values for the same output mRNA given the same input protein. It also occasionally reports an MFE value that does not match the MFE of the sequence as calculated by ViennaRNA \cite{lorenz2011viennarna}. We observe that all software packages target parity with the RNAfold program running with the \texttt{-d0} option.

LinearDesign also exhibited bugs such as sometimes crashing during execution when it produced an invalid RNA sequence for the input protein triggering an assertion error. In some cases LinearDesign produced an mRNA sequence with a reported MFE value that did not match ViennaRNA's computed MFE value. Undefined behavior is the suspected cause of this. We recompiled LinearDesign with sanitization (via \texttt{-fsanitize=address,undefined}). Several cases of integer underflow were detected.

No bugs were observed in CDSfold.

Proteins that trigger the bugs mentioned here are compiled in our GitHub repository. The errors can be reproduced using our code that runs the various software packages or by calling the software packages directly using the same codon usage table and program settings.

\section{Discussion}

The success of LinearDesign and mRNA technology more generally highlights the importance of mRNA folding algorithms. They are fast enough to use for long proteins and provide a high degree of optimization for stability (via MFE) and codon choice (via CAI). However, the lack of flexibility and features is a limitation. In addition, existing software packages are imperfect with the user needing to use different software packages to access different features and contend with bugs.

We have identified several gaps in the mRNA folding literature and also in the available software. Perhaps the most pressing research gap is to incorporate sequence and structure constraints, as these are widely used in existing mRNA optimization approaches. Existing mRNA folding algorithms can still be used to generate an initial sequence, which may be adjusted by another algorithm to meet sequence and structure constraints. However, a holistic approach that can incorporate some of these constraints into mRNA folding is preferred.

Another pressing gap for mRNA folding algorithms is the lack of high-quality software packages. Existing software either have significant bugs (DERNA and LinearDesign), poor performance (DERNA), high memory usage (DERNA and LinearDesign) or lack features (CDSfold and LinearDesign). We also note that no multi-core or GPU-enabled software exists despite the significant computational bottlenecks in mRNA folding algorithms.

There are also several specific ideas that we suggest for the next iteration of mRNA folding algorithms.

\textit{Inclusion of UTRs.} We observe for completeness that it is possible to extend mRNA folding algorithms to be UTR-aware. The UTRs can be incorporated by modifying the codon graph construction without any changes to the recursions. Construct a path for the 5' UTR and the 3' UTR. Each path contains the sequence of nucleotides in the UTR. The 5' UTR path can be prepended to the codon graph and the 3' UTR can be appended. The edges in the UTR paths should have weight zero so that they do not contribute to CAI. This is sufficient to ensure that the UTRs are included in the calculation of the MFE. To our knowledge, the addition of UTRs has not been implemented in existing mRNA folding software packages.

\textit{Suboptimal folding.} Current mRNA folding algorithms only return a single solution, but it would be more practical to provide the user with a diverse set of potential sequences. Suboptimal sampling is one of the most important features of modern RNA folding software packages \cite{mathews2006revolutions,wuchty1999complete,ding2003statistical}. An mRNA folding implementation would mitigate issues with sequence and structure constraints since a diverse set of mRNAs is more likely to contain valid solutions. In addition, it gives a larger pool of potential sequences for lab testing. DERNA finds a set of Pareto optimal solutions \cite{gu2024derna}. This set equivalent to that obtained by running conventional mRNA folding with all $\lambda$ values. It is important to understand how this is different from suboptimal sampling. DERNA finds only a single solution for a given $\lambda$, but there may be many near-optimal solutions. Further, there may be ties for Pareto optimal solutions in which case DERNA will only report one.

\textit{Forbidden sequence avoidance.} There is currently no computer algorithm for building a codon graph (or DFA) for an arbitrary set of sequence motifs to avoid. Though, Zhang \textit{et al.} \cite{zhang2023algorithm} give a bespoke construction for a specific sequence. We hypothesize that this could be achieved by combining the codon graph construction (or DFA) with the Aho-Corasick automaton \cite{aho1975efficient}. In addition, no method has been proposed to avoid repeated sequences or inverted repeats.

\section{Closing Remarks}
mRNA technology is at the cutting edge of therapeutics offering better vaccines, gene-editing, and personalized medicines. mRNA sequences optimization is essential to fully realize this potential, and mRNA folding algorithms are one of the most powerful tools available.

This review explores mRNA folding algorithms, which although recently popularized by LinearDesign, have existed since the early 2000s. We provide a comprehensive description of how these algorithms work, addressing the lack of comprehensive explanation of the core algorithms in literature. Further, we unify and simplify the description of the algorithms used in CDSfold and LinearDesign with a new codon graph framework. Several key gaps in the literature are highlighted and we present benchmarks comparing run-time speed, memory usage, correctness, and features of existing software.

We hope this review provides a strong foundation for the development of next-generation mRNA folding algorithms and contributes to the continued advancement of mRNA technology.

\section{Acknowledgements}
We thank The University of Western Australia and Moderna therapeutics for providing computational resources and support. We are also grateful to Haining Lin and Wade Davis for their valuable discussions and feedback, which helped shape this work. Additionally, we acknowledge the developers of LinearDesign, CDSfold, and DERNA for making their software publicly available, enabling our benchmarking studies.

\bibliographystyle{abbrv}
\bibliography{main}

\clearpage
\section{Supplementary Material}

\begin{figure}[h]
    \centering
    \includegraphics[width=0.8\textwidth]{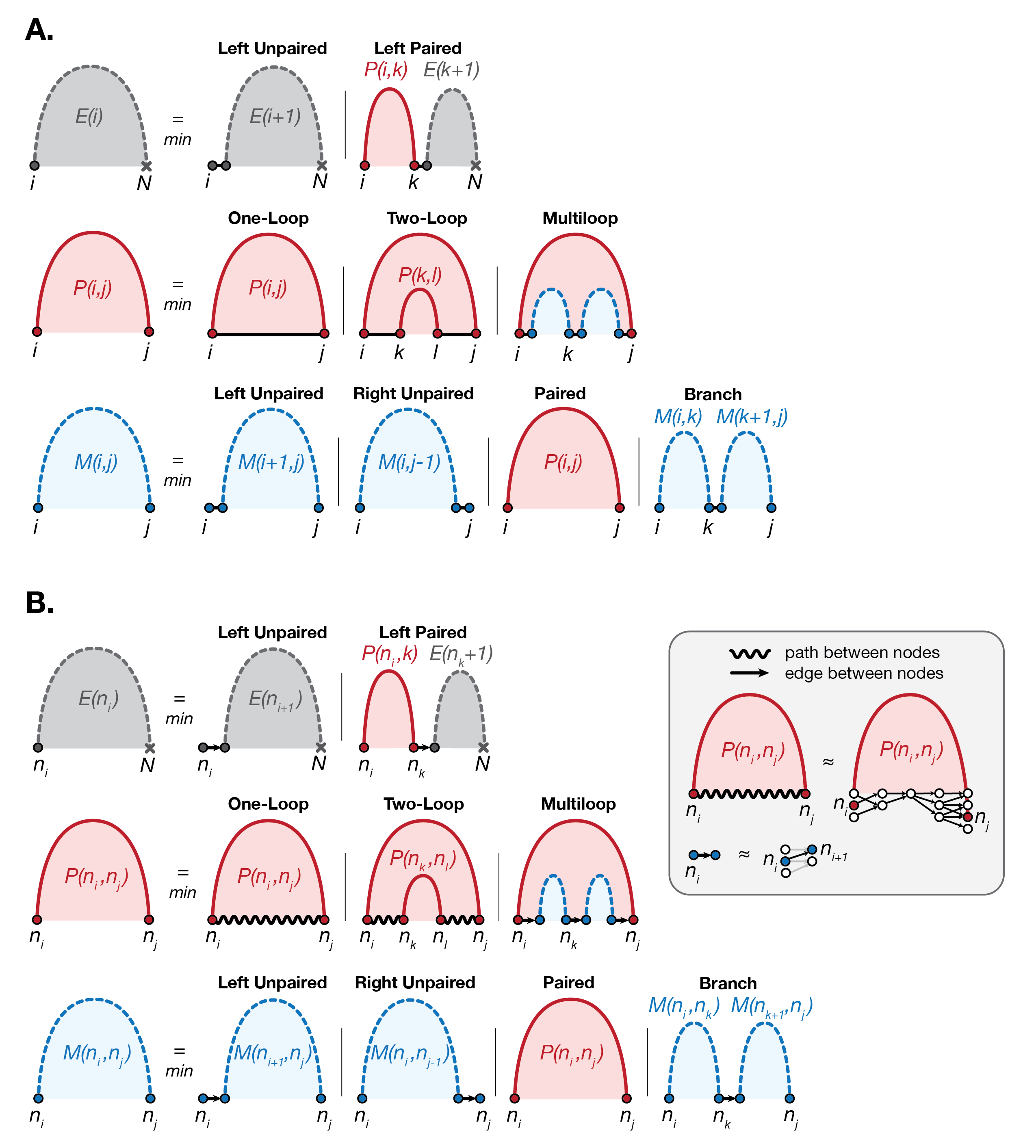}
    \caption{RNA Folding Recursions with Example Structures. (A) represents the Zuker-Stiegler recursions for the external loop function (grey), paired function (red), and multiloop function (blue). Each case of the recursions is depicted with a Feynman-style diagram, and an example RNA substructure corresponding to this diagram. A wavy black line in the Feynman diagram represents a path through the sequence of nucleotides between two positions, while a black arrow indicates that two nucleotides are adjacent in the sequence. (B) compares RNA folding and mRNA folding. The diagrams in (A) can be used to describe both RNA folding and mRNA folding, with a few key differences. In RNA folding, a path between positions is equivalent to the nucleotide sequence between those two indices, since there is only a single sequence. In mRNA folding, indices in the sequence are replaced with pointers to nodes at those same positions, and a path between nodes becomes a path through the codon graph connecting the two nodes. Each path corresponds to a different possible nucleotide sequence. Similarly, an arrow between adjacent positions in RNA folding simply represents a step along the sequence backbone, while an arrow between adjacent nodes in mRNA folding represents a specific edge in the codon graph.}
    \label{fig:figS1}
\end{figure}

\begin{figure}[h]
    \centering
    \includegraphics[width=0.8\textwidth]{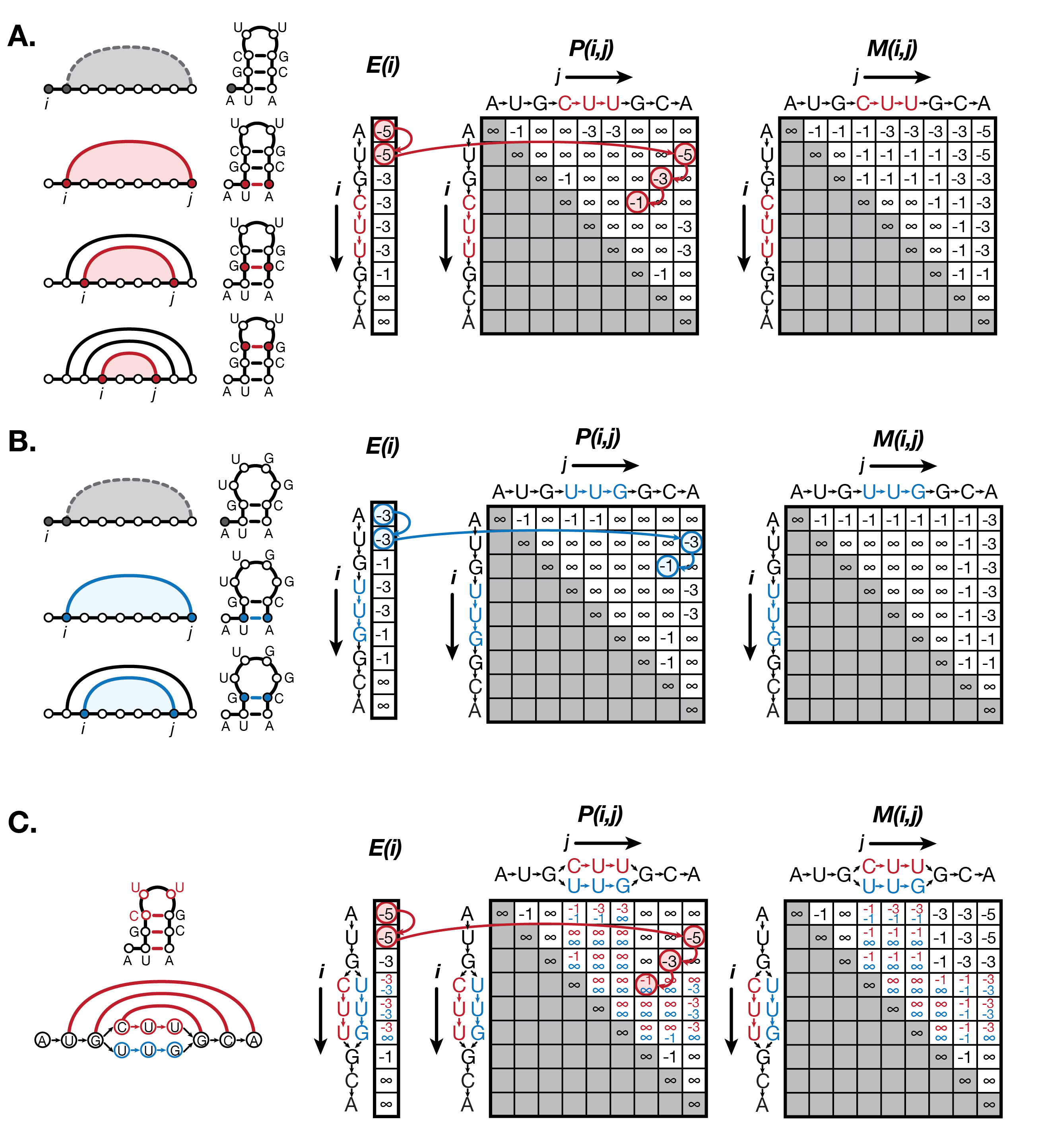}
    \caption{Dynamic Programming for RNA and mRNA Folding. (A) and (B) represents the external loop $E(i)$, paired $P(i,j)$, and multiloop $M(i,j)$ tables for two simple mRNA sequences that code for the same protein. The tables are filled according to the Zuker-Stiegler recursions with a set of simplified energy terms purely for demonstration purposes: we set the energy of a $\textsc{OneLoop}(i,j)$ to -1 kcal, the energy of a $\textsc{TwoLoop}(i,j,k,l)$ to -2 kcal, and the energy of initializing a multiloop to -3 kcal. Note that in these simple examples, we never use the multiloop energy term since the lowest energy structure includes only a stem-loop. The traceback of the lowest energy structure follows the red (A) or blue (B) arrows, beginning at $E(1)$. The arc diagram and secondary structure diagram corresponding to each arrow of the traceback are shown to the left of each respective $E(i)$ table. The two synonymous mRNA sequences from (A) and (B) are combined into a codon graph in (C). The corresponding external loop, paired, and multiloop tables now include energies for both the red and blue codons. The lowest energy structure in this case uses the CUU codon (red) rather than the UUG codon (blue). The traceback follows this red path, corresponding to the structure of the first sequence. }
    \label{fig:figS2}
\end{figure}

\begin{algorithm}
\caption{Traceback algorithm}
\label{alg:traceback}
\begin{algorithmic}[1]
\Procedure{Traceback}{$\alpha$}
    \State \textbf{Input:} An amino acid sequence $\alpha$
    \State \textbf{Output:} A optimized mRNA sequence
    \State Precompute the dynamic programming tables $P$, $M$, and $E$ for the input protein $\alpha$
    \State $S$ = $\{E(1)\}$ \Comment{Stack of table states. Initial state is $E(1)$}
    \State $m \gets \emptyset$ \Comment{The mRNA sequence to output}
    \While{$|S| > 0$} \Comment{Process states until stack is empty}
        \State $s \gets \textsc{Pop}(S)$
        \If{\textsc{IsBaseCase}(s)}
            \State \textbf{continue} \Comment{Skip base cases}
        \ElsIf{$\textsc{Table}(s) = P$} \Comment{Paired table}
            \State $n_i, n_j = \textsc{Params}(s)$
            \If{$\textsc{Value}(s) = \textsc{OneLoop}(b_{n_i}, b_{n_j}) + \text{LCAI}(n_i \leadsto n_j)$} \Comment{OneLoop case}
                \State $\textsc{AddShortestPath}(n_i \leadsto n_j, m)$
                \State \textbf{continue}
            \EndIf
            
            \For{$k : i < k < j$} \Comment{TwoLoop case}
                \For{$l : k < l < j)$}
                    \For{$n_k \in \operatorname{atpos}(k) : R(n_i, n_k)$}
                        \For{$n_l \in \operatorname{atpos}(l) : R(n_l, n_j)$}
                            \State $v \gets \textsc{TwoLoop}(b_{n_i}, b_{n_j}, b_{n_k}, b_{n_l}) + P(n_k, n_l) +  \text{LCAI}(n_i \leadsto n_k) + \text{LCAI}(n_l \leadsto n_j)$
                            \If{$\textsc{Value}(s) = v$}
                                \State $\textsc{AddShortestPath}(n_i \leadsto n_k, m)$ \Comment{Adds bases on the shortest path}
                                \State $\textsc{AddShortestPath}(n_l \leadsto n_j, m)$
                                \State $\textsc{Push}(P(n_k, n_l), S)$
                                \State \textbf{break} nested for-loops and \textbf{continue} while-loop
                            \EndIf
                        \EndFor
                    \EndFor
                \EndFor
            \EndFor
            
            \For{$k : i< k < j$} \Comment{Multiloop case}
                \For{$n_k \in \operatorname{atpos}(k)$}
                    \For{$n_{k+1} \in \operatorname{out}(n_k)$}
                        \For{$n_{i+1} \in \operatorname{out}(n_i)$}
                            \For{$n_{j-1} \in \operatorname{in}(n_j)$}
                                \State $v \gets M(n_{i+1}, n_k) + M(n_{k+1}, n_{j-1}) + \textsc{ML}_{\text{init}} + \textsc{ML}_{p}$
                                \State $v \gets v + \text{LCAI}(n_i, \leadsto n_{i+1}) + \textsc{LCAI}(n_k \leadsto n_{k+1}) + \text{LCAI}(n_{j-1}, \leadsto n_j)$
                                \If{$\textsc{Value}(s) = v$}
                                    \State $\textsc{Push}(M(n_{i+1}, n_k), S)$
                                    \State $\textsc{Push}(M(n_{k+1}, n_{j-1}), S)$
                                    \State \textbf{break} nested for-loops and \textbf{continue} while-loop
                                \EndIf
                            \EndFor
                        \EndFor
                    \EndFor
                \EndFor
            \EndFor
        
        \ElsIf{$\textsc{Table}(s) = M$} \Comment{Multiloop table}
            \State $n_i, n_j = \textsc{Params}(s)$
            \For{$n_{i+1} \in \text{out}(n_i)$} \Comment{Unpaired at $n_i$ case}
                \State $v \gets M(n_{i+1}, n_j) + \textsc{ML}_{u} + \text{LCAI}(n_i \leadsto n_{i+1})$
                \If{$\textsc{Value}(s) = v$}
                    \State $\textsc{Push}(M(n_{i+1}, n_j), S)$
                    \State $\textsc{AddBase}(b_{n_i}, m)$
                    \State \textbf{break} for-loop and \textbf{continue} while-loop
                \EndIf
            \EndFor
            
            \State $\dots$ continued on next page
        \algstore{traceback}
\end{algorithmic}
\end{algorithm}

\begin{algorithm*}
\begin{algorithmic}[1]
\algrestore{traceback}

            \For{$n_{j-1} \in \text{in}(n_j)$} \Comment{Unpaired at $n_j$ case}
                \State $v \gets M(u, n_{j-1}) + \textsc{ML}_{u} + \text{LCAI}(n_{j-1} \leadsto n_j)$
                \If{$\textsc{Value}(s) = v$}
                    \State $\textsc{Push}(M(n_i, n_{j-1}), S)$
                    \State $\textsc{AddBase}(b_{n_j}, m)$
                    \State \textbf{break} for-loop and \textbf{continue} while-loop
                \EndIf
            \EndFor

            \If{$\textsc{Value}(s) = P(n_i,n_j) + \text{ML}_{p}$} \Comment{Branch case}
                \State $\textsc{Push}(P(n_i,n_j), S)$
                \State \textbf{continue}
            \EndIf

            \For{$k : i < k < j$} \Comment{Bifurcation case}
                \For{$n_k \in \operatorname{atpos}(k)$}
                    \For{$n_{k+1} \in \operatorname{out}(n_k)$}
                        \State $v \gets M(n_i, n_k) + M(n_{k+1}, n_j) + \text{LCAI}(n_k \leadsto n_{k+1})$
                        \If{$\textsc{Value}(s) = v$}
                            \State $\textsc{Push}(M(n_i, n_k), S)$
                            \State $\textsc{Push}(M(n_{k+1}, n_j), S)$
                            \State \textbf{break} nested for-loops and \textbf{continue} while-loop
                        \EndIf
                    \EndFor
                \EndFor
            \EndFor
        
        \Else \Comment{External loop table}
            \State $n_i = \textsc{Params}(s)$
            \For{$n_{i+1} \in \text{out}(n_i)$} \Comment{Unpaired at $n_i$ case}
                \If{$\textsc{Value}(s) = E(n_{i+1}) + \text{LCAI}(n_i \leadsto n_{i+1})$}
                    \State $\textsc{Push}(E(n_{i+1}), S)$
                    \State $\textsc{AddBase}(b_{n_i}, m)$
                    \State \textbf{break} for-loop and \textbf{continue} while-loop
                \EndIf
            \EndFor
            
            \For{$k : i < k < N$} \Comment{Branch case}
                \For{$n_k \in \operatorname{atpos}(k)$}
                    \For{$n_{k+1} \in \operatorname{out}(n_k)$}
                        \If{$\textsc{Value}(s) = P(n_i, n_k) + E(n_{k+1}) + \text{LCAI}(n_k \leadsto n_{k+1})$}
                            \State $\textsc{Push}(P(n_i, n_k), S)$
                            \State $\textsc{Push}(E(n_{k+1}), S)$
                            \State \textbf{break} nested for-loops and \textbf{continue} while-loop
                        \EndIf
                    \EndFor
                \EndFor
            \EndFor
        \EndIf
    \EndWhile
    \State \textbf{return} $m$
\EndProcedure
\end{algorithmic}
\end{algorithm*}

\end{document}